# Data-Folding and Hyperspace Coding for Multi-Dimensional Time-Series Data Imaging

## Author Information

**School of Information Science and Engineering, Northeastern University, Shenyang, China**

Chao Lian, Yuliang Zhao

**School of Electrical Engineering, Yanshan university, Qinhuangdao, China.**

Zhikun Zhan

**CAS-CityU Joint Laboratory for Robotic Research, City University of Hong Kong, Hong Kong, SAR, China**

Wen J. Li

## Abstract

  Multi-Dimensional time series classification and prediction has been widely used in many fields, such as disease prevention, fault diagnosis and action recognition. However, the traditional method needs manual intervention and inference, and cannot realize the figurative expression of multi-Dimensional data, which lead to inadequate information mining. Inspired by the strong power of deep learning technology in image processing, we propose a unified time-series image fusion framework to transform multi-modal data into 2D-image, and then realize automatic feature extraction and classification based on a lightweight convolutional neural network. We present two basic image coding methods, Gray image coding, RGB image coding, and their step coding methods. Considering the universality of different application fields, we extended the coding method and propose two types transform coding, Transform-RGB coding and RGB-Transform coding, to improve the multi-domain representation ability. By applying to three typical scenes of Parkinson's disease diagnosis, bearing fault detection and gymnastics action recognition, we obtained the highest classification accuracy of 100%, 92.86% and 99.70% respectively, which were all higher than the classical processing methods. It proves the strong classification ability and universality of our coding framework to different multi-dimensional scenes. We expect that this method can be used and perform well in other scenarios, and be potential to facilitate the progress of related technology.

## Introduction

Sensor sensing technology is one of the three pillars of modern information industry and an important milestone in the evolution of science and technology. Real-world applications such as motion recognition[1,2], fault diagnosis [3,4,5], and medical care [6,7] can produce multi-dimensional time-series data that requires processing. However, this problem has not be well solved due to the lack of a universal cross-domain fusion framework for such multi-modal sensor data. In general, the processing methods of sensor time series data can be divided into model-based, signal-based, knowledge and hybrid active-based methods. The knowledge-based approach is usually based on a large number of data driven, which does not need a priori model and signal generation structure, only need to realize the deconstruction of the state on the basis of data mining. In recent years, with the rapid development of sensor technology and big data technology, more data are collected, which also provides opportunities for data-driven information mining.

In this context, data mining models based on expert knowledge and machine learning become mainstream. Some scholars achieve the recognition task by obtaining internal features of the original time series data, such as statistical features, frequency domain features, and other possible feature forms, based on expert knowledge. For example, in the field of fault diagnosis, some scholars use priori data feature experience to extract frequency domain features for analysis and decoding; In the field of behavior recognition, scholars tend to define different behavior patterns by introducing artificial statistical features, such as amplitude, maximum value, and root mean square, etc. With the development of deep learning technology, some scholars have tried to apply deep learning technology to time series data to replace the process involved by expert features [11,10,8,9]. For example, in the medical field, some scholars have introduced multi-channel convolutional models to achieve the diagnosis of heart rate diseases, sleep stages, etc.; In the field of fault diagnosis, scholars introduced one-dimensional convolutional neural networks to monitor machine failures, covering different application models and improved models.

However, there still two main problems need to be solved. First, for different application fields, there are big differences in the feature extraction methods and model frameworks, and people still need to select and improve the model according to specific scenarios; Second, the advantages of deep learning networks have not been really exerted, and their overall improvement effect is limited. Therefore, how to give greater play to the advantages of deep learning technology in time series data processing has become an urgent problem to be solved.

In order to solve the above problems, some scholars try to encode time series data into images, and then better use the powerful ability of deep learning technology in image recognition [12][13][14]. In this way, the feature region of the sequence can be enlarged and the temporal correlation can be constructed, thus improving the accuracy. For example, some scholars try to use GAF image [15][16][17], MTF image, RP image and other ways to encode time series data, so as to realize data mining through deep learning technology [18][19]. There are also some scholars use multiple image coding methods to realize data mining. To a certain extent, this method effectively makes use of the powerful advantages of deep learning in image processing, and has achieved success in finance, medical treatment, industry and other fields. In addition, in specific fields, there are also some special encoding methods are used for data mining. For example, in the field of fault diagnosis, scholars try to transform the original data into SIFT images and spectrum images for information mining.

However, although these approaches have made great contributions, there are still some urgently problems need to be solved. First, these methods are mainly aimed at low-dimensional data, especially one-dimensional data, which limits its ability to apply in other complex environments. Though some scholars attempt to develop multi-channel deep learning network for processing, it lead to more complex computation and large model; Second, the application scenario capabilities of this method are limited, such as SIFT images and spectrograms, which are more suitable for fault diagnosis and other similar fields. Thus, there is a lack of unified data mining mode that adapts to the needs of multi-scenario applications; Third, the data dimension of images obtained with these encoding methods will be massively increased, especially when the data is multi-dimensional.

To solve the above problems, we propose a comprehensive time series data imaging framework, *Data-Folding and Hyperspace Coding Method* (DFHC) for multi-dimensional time-series data imaging. This framework takes into account the needs of different application scenarios and covers all aspects from image coding, image extension to deep learning network models. Specifically, we propose two basic coding methods, grayscale coding and RGB encoding. On this basis, we propose the step images, RGB-Step and Gray-Step images to reflect the ability of locally changing information. To meet the needs of different application scenarios, we further proposed two type of extended coding methods, namely TF-RGB and RGB-TF for the improvement of the multi-domain information expression capability. Finally, we develop an improved LeNet 5 convolutional neural network model to achieve fast and accurate data mining.

## Results

Experiments were developed with three typical scenarios, including fault diagnosis, medical monitoring, and behavior recognition, to verify the effectiveness and universality of our method. According to different scenes, we use basic coding image, Step image and extended image to carry on the detailed demonstration.

### Intelligent industry area: Fault diagnosis

In the field of intelligent industry, rolling bearing failure is a common phenomenon. Rolling bearing is an important transmission structure that directly affects the transmission efficiency and safety of rotating machinery equipment. It is also the most frequently used and the most seriously worn part in machinery equipment. Therefore, bearing condition monitoring and fault diagnosis are of great significance to improve the safety of rotating machinery.

This part adopts the motor bearing fault dataset published by Case Western Reserve University in the United States. The rolling bearing model used in the experiment is 6205-2RS JEM SKF. The bearing fault signal sampling test bench is shown in Fig. 1 (a). In this dataset, there are three main fault types, including rolling body fault, outer ring fault and inner ring fault. Each fault type has three different damage sizes, which are 0.18 mm, 0.36 mm and 0.54mm, respectively. Therefore, a total of 10 states can be finally obtained, including 9 fault states and 1 normal state. In this dataset, two channels acceleration data of the drive end (DE) and the fan end (FE) are collected. The sampling frequency of the data is 12K.

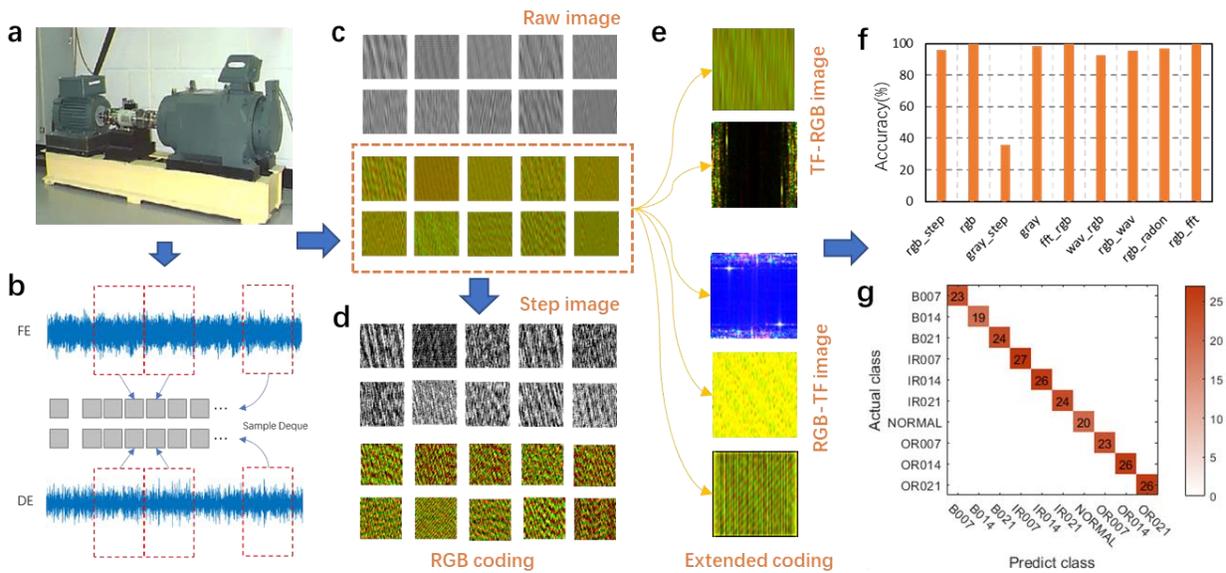

**Fig. 1 Demonstration of fault diagnosis.** a Schematic diagram of experimental method. b coding method of RGB image for Fault diagnosis dataset. c basic coding images, include Gray image and RGB image for 10 states of motor bearing. d step images, include Gray-Step image and RGB-Step image for 10 states of motor bearing. e extended coding images, include RGB-TF image and TF-RGB image for 10 states of motor bearing. For TF-RGB image, there are FFT-RGB and WT-RGB image, and for RGB-TF image, there are RGB-FFT, RGB-Radon and RGB-WT image. f fault diagnosis results with different coding methods. g confusion matrix of RGB-FFT method.

Fig 1 (b) shows our coding method for fault data. We obtain each time series segment by window interception, and for each adjacent window, there is no overlap between them. Considering that there are only FE and DE channels, we reflect FE and DE channels correspond to R channel and G channel respectively, and B channel is set to 0. For gray images, there is nothing special process beside direct stacking. As shown in Fig. 1 (c), for different fault state images. this difference is reflected in texture and color differences for color images, while it is reflected in texture and brightness for grayscale images. As shown in Fig. 1 (d), the obtained Step image also has distinct features, but the overall it is worse than the original image, because of the rapid vibration in fault signal. Therefore, the internal small waveform change information has little contribution to the classification work. As shown in Fig. 1 (e), we see very clear style differences for the extended coding images, which means they have great differences in the perspective of reflecting information, and this characteristic may lead to different classification performance, as shown in Fig. 1 (f).

In the experiments, according to the characteristics of fault data, we convert all color and gray images obtained into 64*64 size. The radio of training set, validation set and test set at 7:1:2. As shown in Fig. 1 (f), the original RGB image has also reached an impressive accuracy at 99.58%. Most surprisingly, in all the encoded images, the frequency domain coding method achieves the best performance. The accuracy of FFT-RGB and RGB-FFT coding even reached at 100%. which is consistent with the consensus that the frequency domain analysis method has a strong ability in fault data diagnosis.

## Medical monitoring area: Parkinson diagnosis

Parkinson's disease (PD) is a chronic neurodegenerative disease in middle-aged and elderly people. Accurate and effective detection of Parkinson's disease is not only a medical problem, but also a social public health problem, which is crucial for later medical intervention and health management. Parkinson's disease patient develop a range of motor and non-motor disorders, highlighted by uncontrolled shaking of their hands and the back of their hands. Therefore, the early diagnosis and recognition of Parkinson's disease can be effectively carried out through handwriting dynamic examination.

This part adopt the Parkinson's diagnosis dataset NewHandPD dataset, which was collected by the Botucatu Faculty of Medicine, National University of Sao Paulo, Brazil. The dataset includes a total of 66 individuals, including 31 PD patients and 35 healthy individuals, ranging in age from 14 to 79 years. In the healthy group, there were 18 males and 17 females, while there were 21 males and 10 females in the patient group. Each individual was asked to perform 12 handwriting dynamics checks using a smartpen. In this paper, we use the left-handed dichotomous motion dataset for experiment.

The smartpen is equipped with six different sensors, and the corresponding six-channel sensor signals are collected when an individual is doing a handwriting check. The position of six sensors of the smart pen are in four different positions, as shown in Fig 2 (a). The first three channel signals come from writing pressure, writing grip force, and ink filling pressure sensors, and the other three channel signals come from tilt and acceleration sensors in the X, Y and Z directions, respectively.

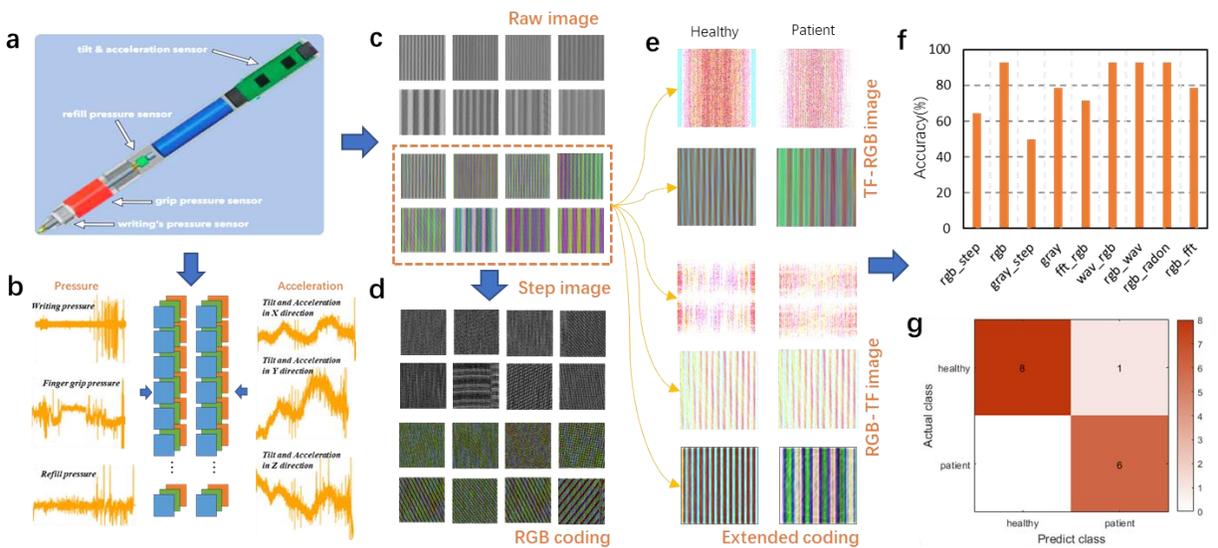

**Fig. 2  Demonstration of Parkinson's disease diagnosis. a** Schematic diagram of Smart Pen. **b** coding method of RGB image for NewHandPD dataset. **c** basic coding images, include Gray image and RGB image. **d~e** represent coding result of healthy and patient group, where the first row represents coding images of healthy group, and the second row represents coding images of patient group. **d** step images, include Gray-Step image and RGB-Step image. **e** extended coding images, include RGB-TF image and TF-RGB image. For TF-RGB image, there are FFT-RGB and WT-RGB image, and for RGB-TF image, there are RGB-FFT, RGB-Radon and RGB-WT image. **f** Parkinson's diagnosis results with different coding methods. **g** confusion matrix of RGB-Radon method.

In the encoding process, we take the data obtained from the pressure sensor as one cluster, and the data from the tilt and acceleration sensors as the other cluster, and then 2D images with only a single line for each cluster can be obtained through RGB coding method, as shown in Fig. 2 (b). By merging the RGB images for two cluster, two lines of RGB queue images can be obtained. For grays image, the images can be obtained by the method of six-channel data stacking, including three signals from pressure sensors and tilt and acceleration sensors. Finally, the final RGB or Gray image can be obtained by folding deduction, as shown in Fig. 2 (c). We can clearly see the difference in image stripes between different types of actions. Furthermore, the RGB-STEP image and Gray-STEP image are obtained by numerical difference of the raw data, in order to obtain the local variation features of signals, as shown in Fig. 2 (d). The obtained extended coding images are shown in Fig. 2 (e), which are divided into Transform-RGB coding (TF-RGB) and RGB-Transform coding(RGB-TF). TF-RGB image includes RGB coding after FFT transformation(RGB-FFT) and wavelet transform(RGB-WT), which are used to establish the potential coding method applied in not only time series signals. RGB-TF coding includes FFT transform, wavelet transform and Radon transform of RGB images, which is used to obtain and study the transform domain features of RGB coding images, so as to extend and explore the multi-scene adaptability of RGB coding. And we can also find the variation in the form of feature expression of different extended coding images.

The dataset in this experiment consisted of 66 subjects, including 35 healthy subjects and 31 patients with Parkinson's disease. The ratio of training set, validation set and testing set is 7:1:2. The final recognition result is shown in Fig. (f). It can be seen that, as a proposed basic coding, RGB coding has a profound effect, with an accuracy rate of 92.86%, while gray coding has a relatively poor effect, with only 78.57%. This may be caused by its insufficient revelation on the relationship between different channels. Meanwhile, WT-RGB, RGB-WT and RGB-Radon all achieve the similar classification performance. Comparatively speaking, the overall classification accuracy in frequency domain coding is low, for example, only 78.57% for RGB-FFT coding. For step coding, the performance is the worst, even it is only 50% for gray coding. The most likely reason is that this type of data belong to long time series high-frequency vibration data, and the features of internal waveform state are not so important for classification.

## Behaviour recognition area: Gymnastics action recognition

The popularity of sports has deepened, and the forms of fitness have gradually shown a trend of diversification. Among them, gymnastics, because of its rich body movements, can not only regulate the physiological function of human body systematically, but also play a positive role in human mental health. As a main part, accurate recognize gymnastic movements can be of great significance for realizing scientific gymnastic exercises.

In this part, we used the gymnastics dataset, which was acquired and published by Northeastern University. This dataset is collected by a wireless gymnastics action recognition system composed of 11 nodes, covering of arm, wrist, thigh and manic joint, etc. as shown in Fig. 3 (a). A total of 7 volunteers were included in the experiment, including 4 man and 3 woman. All the volunteers were asked to perform 6 types of gymnastics, including stretching, chest expansion, body side, body turn, kick and whole body. The data were completed in 5 experiments. In each experiment, the volunteers were asked to repeat each action 8 times, and finally for each type of gymnastic, there are totally 40 samples for each volunteer. The data acquisition of each sensor node is completed by the MPU6500 sensor, which is integrated with the 3-axis accelerometer and 3-axis gyroscope. The sampling frequency is set at 50 Hz. In this paper, only the data of the left arm are used.

The coding method here is similar to the previous case, we divided the 6 channels into two clusters, acceleration and angular velocity, respectively. For each cluster, RGB coding is carried out to obtain the final RGB image, and then be stacking and resized, as shown in Fig. 3 (b). Further, we obtain the corresponding Gray image and RGB image, as shown in Fig (c). As can be seen, there is a certain similarity between the two images. For grayscale images, the main difference between categories lies in the brightness and texture, while RGB images are reflected in the color depth and texture of. Step image is similar to the classic two-dimensional code and color two-dimensional code, and the feature expression is slightly insufficient, as shown in Fig. 3 (d). Fig. 3 (e) shows the results of different extended coding images, and there are great differences in the representation of different coding methods.

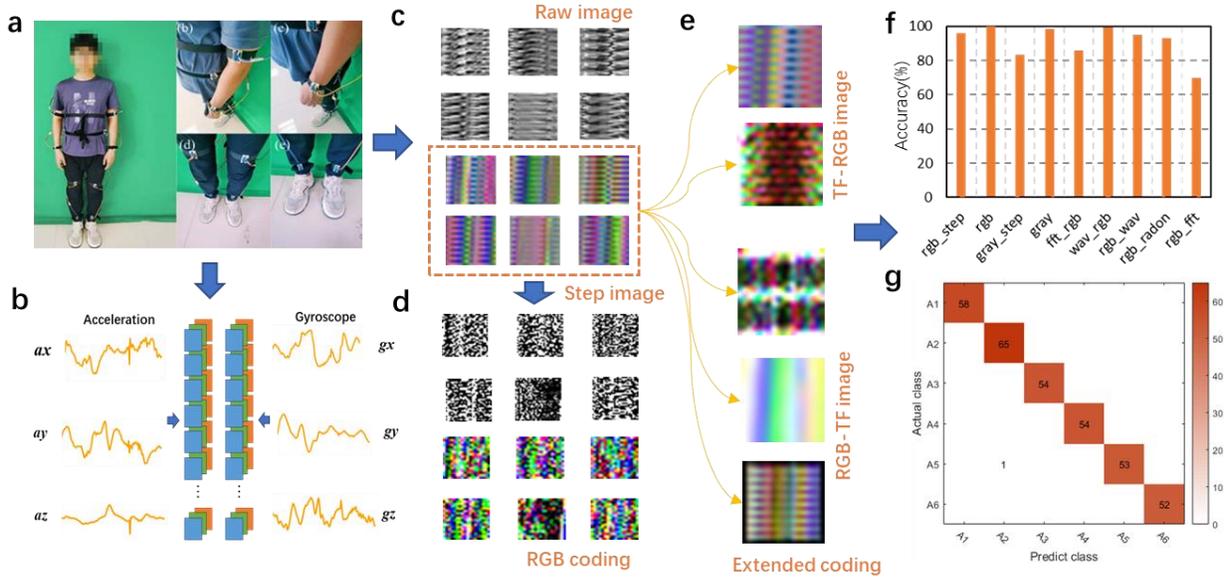

**Fig. 3 Demonstration of Gymnastics action recognition. a** Schematic diagram of Gymnastic data acquisition system. **b** coding method of RGB image for Gymnastics dataset. **c** basic coding images for 6 types of gymnastics, include Gray image and RGB image. **d** step images for 6 types of gymnastics, include Gray-Step image and RGB-Step image. **e** extended coding images for 6 types of gymnastics, include RGB-TF image and TF-RGB image. For TF-RGB image, there are FFT-RGB and WT-RGB image, and for RGB-TF image, there are RGB-FFT, RGB-Radon and RGB-WT image. **f** Gymnastics action recognition results with different coding methods. **g** confusion matrix of RGB-WT method.

According to the previous description, we ended up with a total sample number of 1680 samples, with 280 samples for each action. The ratio of training set, validation set and test set is 7:1:2 in the experiment. As shown in Fig. 3 (f), Gray coding, RGB coding and WT-RGB coding have achieved good recognition results, with accuracy of 98.51%, 99.70% and 99.40%, respectively. The accuracy of RGB coding and WT-RGB coding is the best. In contrast, the feature expression of frequency domain coding is poor, and the highest is only 86.01%. Therefore, it can be speculated that frequency domain features may not be suitable for action recognition applications.

## Discussion

Multidimensional time series data are widely used in current production and life. Among them, the common time series data mostly appear in the form of clusters, such as acceleration data, which has 3 directions, and the same as 3-axis angular velocity data and magnetic field data. Other data with similar types, such as position coordinate data (*x, y, z*) can be also mapped to RGB images in the same way. Through RGB coding method, correlations between different channels of these data can been revealed at a deeper level and the raw time series signal can be better analyzed with deep learning algorithms. In particular, the coding method can be extended to other scenarios with similar relationships, such as two-channel acceleration data in fault diagnosis and multidimensional pressure data in Parkinson's diagnosis studied in this paper. These time series data all occur at the same time, so the representation method in this paper can be naturally used as well.

In addition to the basic coding methods, this paper also proposes a variety of types of extended coding methods to adapt to the specific coding requirements of different scenarios, which is also confirmed by the experimental results. As shown in Table 1, we summarize the recognition results of different coding methods in three scenarios.

**Table 1 Summary of the recognition results of various coding methods in three typical scenarios.**

| No | Coding method | Method | Recognition accuracy (%) | | |
|---|---|---|---|---|---|
| | | | Fault diagnosis | Parkinson diagnosis | Gymnastics recognition |
| 1 | Baseline | -- | 99.43 | 72.56[20] | 97.89 |
| 2 | Raw image coding | Gray | 99.00 | 78.57 | **98.51** |
| 3 | | RGB | **99.58** | **92.86** | **99.70** |
| 4 | Step coding | RGB-Step | 96.20 | 64.29 | 95.83 |
| 5 | | Gray-Step | 35.79 | 50.00 | 83.33 |
| 6 | TF-RGB coding | FFT-RGB | **100.00** | 71.43 | 86.01 |
| 7 | | WT-RGB | 92.41 | **92.86** | **99.40** |
| 8 | | RGB-WT | 95.36 | **92.86** | 95.24 |
| 9 | RGB-TF coding | RGB-Radon | 97.05 | **92.86** | 93.15 |
| 10 | | RGB-FFT | **100.00** | 78.57 | 69.94 |

We can clearly see that, among the basic coding methods, both gray coding and RGB coding can achieve better results, especially RGB coding, which has higher recognition accuracy. This may because it takes more into account the correlation between time series data of the same mode. In the extended coding, we can regard it as two kinds. One is the ordinary transformation coding, such as Step coding, WT-RGB and RGB-Radon. These coding methods have a higher recognition accuracy than that in fault diagnosis and Parkinson's recognition. The reason may be that these ordinary transformation coding may be more suitable for such low-frequency scenarios with obvious waveform variation characteristics. However, the frequency domain coding, such as FFT-RGB and RGB-FFT, has a higher recognition performance in the field of fault diagnosis, which also confirms that extended image in the frequency domain is better suited to this kind of high-frequency vibration scene. This conclusion is consistent with the common-sense of time series processing.

From the perspective of recognition accuracy, the whole experimental process also confirms the strong classification ability and universality of our coding framework to multi-dimensional scenes. By comparing with the basic reference, our method can achieve higher accuracy. On the whole, RGB coding often achieves high classification accuracy, which can be seen as a general coding method. And Gray coding method can be also seen as a general method. Besides, for different application scenarios, a better method is to select targeted coding method with full consideration of the characteristics of the scene, and thus obtain better accuracy.

## Methods

***Data preparation*:** In this study, we use the authoritative data sets of three typical areas to test the broad applicability of our method. The experimental data set contains a fault diagnosis data set from Case Western Reserve University, a data set of Parkinson's diagnosis from National University of Sao Paulo and the gymnastics behavior recognition data set from Northeastern University. Considering the characteristics and recognition requirements of different data sets, we use sliding windows to obtain the entire data set for the fault diagnosis and the diagnosis data set, while for the gymnastics action recognition data set , we just get the activity segment from data set after extraction. Finally, we can obtain the experiment data sets in this study, that is, the acceleration data of the two bits (DE and FE) for the fault diagnosis data set, the pressure data of the three sensors and the acceleration data in the three directions for the diagnosis data set, 3 direction acceleration data and 3 direction angular velocity data for the gymnastics action recognition data set.

*Time-series data imaging method:* Before encoding the data, it is necessary to clarify the object of study in this article. We call the collection of data series with a certain correlation between multiple dimensions as a cluster, which can be a representation of multiple angles of a thing, such as the components of acceleration data in different directions, or it can be a representation of the different spatial locations of a thing, such as the data components of fault events in different locations. In addition, the data can be multimodal, that is, multiple clusters.

For arbitrary data, set the number of clusters to $c$, and each cluster contains a set of components of data of $q$ ($q_1$, $q_2$, …, $q_c$), where $1 \leqslant q_i \leqslant 3$. The reason why it is set to 3 is that the dimensions of the same type of data are currently 3 dimensions at most in practical applications, such as coordinate data, acceleration, angular velocity, magnetic field, Euler angle, etc. However, the number of clusters is not limited.

For each dimension of data under each cluster, in order to ensure that the dimensions of the transformed data are consistent, it is necessary to normalize it to a maximum and a minimum. The formula can be expressed as equation(1):

$$X(t) = \frac{S(t) - \min(S)}{\max(S) - \min(S)} \tag{1}$$

where $S$ represents the original data and $X$ is the converted data.

*(1) RGB image coding:* For any cluster, set the data of different dimensions after conversion are $x$ ($x_1$, $x_2$, …, $x_L$), $y$ ($y_1$, $y_2$, …, $y_L$), $z$ ($z_1$, $z_2$, …, $z_L$), we can establish their correspondence with the color channels R, G, B, that is

$$R: I(1,:,1) = \{x_1, x_2, ..., x_L\}$$
$$G: I(1,:,2) = \{y_1, y_2, ..., y_L\}$$
$$B: I(1,:,3) = \{z_1, z_2, ..., z_L\} \quad (2)$$

With this method, we could obtain a single line of colorful images. And the same operation could also be conducted with data in other cluster. Then we get the original RGB image, which has the dimensions of $n*c$. In order to better play the advantages of convolutional neural networks, we reshape them to make them square pictures. The method can be shown as follows:

Let the total length of valid data be $l$ and the true length be $L$. For the data with $c$ clusters, the total amount of data needed to generate an image is $\sum_{i=1}^{c} l \times q_c$. Considering that the color image needs to be three-channel, the data with dimension less than 3 in the cluster is zeroed out. Let the number of pixels in the converted picture be $p$, then

$$p = l \times c \quad (3)$$

Similarly, according to the formula for the relationship between the width $w$ and pixels $p$ after conversion, we can get

$$p = w \times w \quad (4)$$

However, in real applications the data length may not be exactly L=w*w. So we can always get the inequality $l < L$. Thus, the preliminary solution formula for $w$ can be written as:

$$w = floor\left(\sqrt{c \times l}\right) \quad (5)$$

Where floor is a downward integral function. Considering that the original data may exist in the form of multiple clusters, $w$ also needs to be an integer multiple of the number of clusters $c$ to guarantee that the folded image is square. Therefore, the width $w$ needs to satisfy the following relation:

$$mod(w/c) = 0 \quad (6)$$

Where *mod* represents the remainder function. When *w* does not satisfy the preceding condition, we subtract *w* until the condition is met. In turn, the required valid data length *l* can be reversed and then down sampled by cubic spline interpolation to obtain the final valid sequence. At this point, the final RGB coding picture with a width of *w* can be obtained. As shown in Fig. 4, we present the RGB coding method.

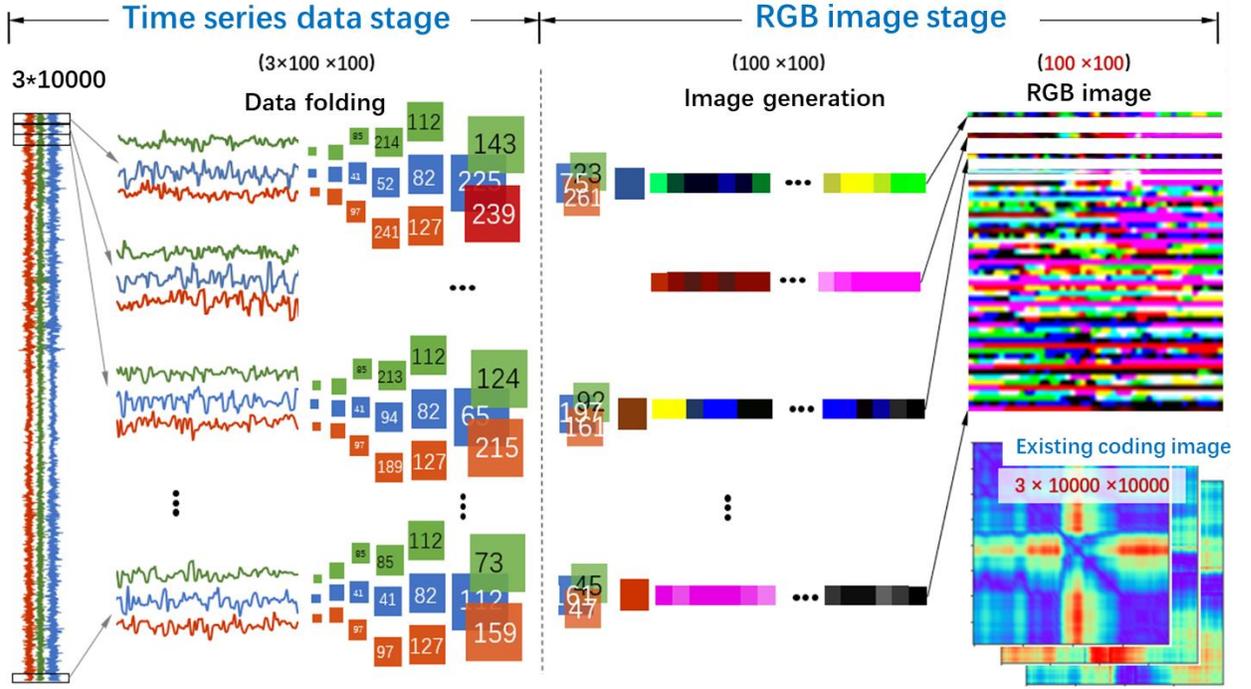

**Fig. 4 Basic RGB coding method. (这里展示本文的整体编码方法图,需要重画)**

*(2) Gray image coding:* For grayscale encoding, we ignore the concept of clusters and reconstruct the data image purely as a channel overlay. Therefore, for the same data, we will get a raw grayscale image with width $\sum_{i=1}^{c} q_c$ and length *l*. Then, in a similar way to the color image, the width of the folded image can be obtained. The calculation formula is:

$$w = floor\left(\sqrt{r \times l}\right) \tag{7}$$

Where *r* represents the total dimension of the original data, which can be expressed as:

$$r = \sum_{i=1}^{c} q_c \tag{8}$$

In addition, the restriction condition of *w* should be changed as:

$$mod(w/r) = 0 \tag{9}$$

*(3) Step coding:* RGB coding and Gray coding solve the global information reflected by the original data. In order to expand the ability of the coding method and reflect the local change information, we further propose the step coding method. This method is designed to process the original data sequence differentially and then carry out RGB coding and Gray coding. For the normalized original data series X(t), the following can be obtained:

$$SX(t) = X(t+s) - X(t), \quad t = 1,...,l-s \tag{10}$$

In the above equation, *s* represents the differential step and *SX(t)* represents the original data series after the difference. Then, the time series after the difference can be encoded as images according to the corresponding formula, that is, RGB-Step encoding and Gray-Step encoding.

*(4) Extended image coding method:* Toward to different applying situation, the coding method can be transformed into different type of forms, such as FFT image, wavelet image and Radon image. According to the order of the transformations, we can divide them into two forms, namely TF-RGB (Transform-RGB encoding) and RGB-TF (RGB-Transform encoding).

*TF-RGB coding：* As the name implies, this method first transforms the original data sequence and then performs RGB coding. This encoding method extends the applicability of the raw RGB encoding method, which is specific to time series data. Here, we extend it to apply to other transform domain types of data. As we all know, in some situations, time series observation cannot observe the internal state information, while transform domain can. Therefore, this paper develops FFT-RGB, WT-RGB these two coding forms.

*Fourier transform* is the basic method of frequency domain signal analysis, it can describe the complex signal better, so it is widely used. For FFT-RGB coding, the transformation formula can be shown below:

$$F(u) = \sum_{t=0}^{T-1} X(t)W^{ut}, u = 0,1,2,...,T-1 \quad (11)$$

Where $W = e^{-j\frac{2\pi}{T}}$. Finally, we get the absolute value of the Fourier transform as final result. Also, we move the zero-frequency component to the center of the spectrum for feature expression.

*Wavelet transform* is used for signal decomposition and reconstruction in this paper. Wavelet decomposition decompositions the original signal into sub-signals of different frequency bands by using scale function $\phi(t)$ and wavelet function $\psi(t)$. The wavelet approximation coefficient $a_0(k)$ and wavelet detail coefficient $d_j(k)$ of a discrete signal $X(t)$ of length $l$ can be expressed as:

$$a_0(k) = \frac{1}{\sqrt{l}} \sum_{i=1}^{l} s(i)\phi_{0,k}(i) \quad (12)$$

$$d_j(k) = \frac{1}{\sqrt{l}} \sum_{i=1}^{l} s(i)\psi_{j,k}(i) \quad (13)$$

Where *j* and *k* represent the expansion and translation of the sub signal in the frequency domain and the translation in the time domain, respectively. The original signal can then be reconstructed according to the approximation coefficient and the detail coefficient.

$$s(t) = \frac{1}{\sqrt{l}} \sum_{k} a_0(k)\phi_{0,k}(t) + \frac{1}{\sqrt{l}} \sum_{j} \sum_{k} d_j(k)\psi_{j,k}(t) \quad (14)$$

When processing using wavelet analysis, this article selects db3 in the Daubechies wavelet function family as the wavelet function.

*RGB-TF coding*: This encoding method is the opposite of the Transform-RGB encoding method, which first encodes RGB image and then uses image transformations to explore more possible feature representations. The purpose of this approach is to explore whether the encoded image has similar transformation domain features to other regular images. In this way, we will be able to further expand the limited time series image information, and then better observe the information mapping of the original data in more dimensions. In this paper, three coding methods of RGB-WT, RGB-Radon and RGB-FFT are mainly applied.

For RGB-FFT and RGB-WT, we applied the 2-D transformation for acquire the final images. The Radon transform is an integral transformation that integrates a function $f(x,y)$ defined in a two-dimensional plane along any straight line on the plane, equivalent to a CT scan of the function $f(x,y)$. Set the pixel of image as $I(x,y)$, then the formula of Randon coding method can be shown below:

$$R(\rho,\theta) = \int_{-\infty}^{\infty}\int_{-\infty}^{\infty} I(x,y)\delta(x\cos\theta + y\sin\theta - \rho)dxdy \tag{15}$$

Where ρ represents the distance from the origin of the coordinates to the projection line L. θ is the angle between the normal line of L and the X-axis, which ranges from 0 to π.

**Architecture of applied CNN model:** In this paper, a new lightweight convolutional neural network is constructed to realize the classification of encoded images, which is an improvement of LeNet5 network to ensure the highest running speed and high accuracy. The raw input of LeNet5 network is 32*32. In the fault diagnosis and Parkinson's diagnosis scenarios in this paper, the size of the image is 64*64 considering of the large volume of signals. In the third scenario, the image size is 32*32 because the waveform length of each segment is relatively small.

The improved network includes three types of layers, namely convolutional layer, pooling layer and FC layer. The convolutional layer extracts the features in the original image through the filter, and the pooling layer is used to reduce the input feature dimension to realize the down-sampling function. After four alternating convolution layers and pooling layers, FC layer is used to calculate the probabilities of each category, and their activation functions are relu and softmax, respectively. In order to control the dimension of feature map and avoid dimension loss, zero filling is adopted in all filling methods. Fig. 5 shows the basic structure of CNN designed in this paper.

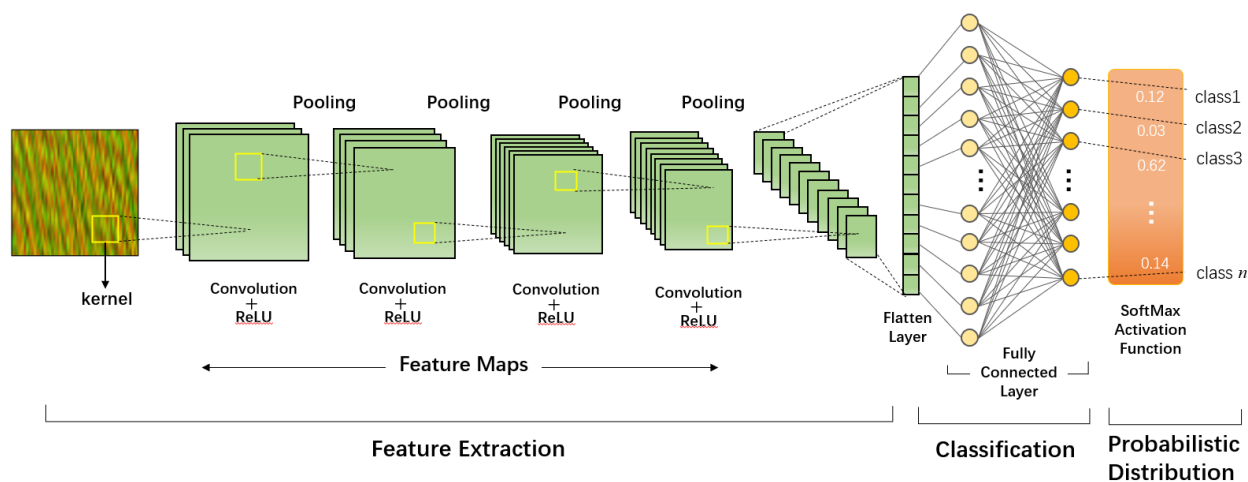

**Fig 5. Proposed CNN structure for time to series images.**

## Data Availability

For more information on data availability, click here. Certain data types must be deposited in an appropriate public structured data depository (details are available here) and the accession number(s) provided in the manuscript. Full access is required at acceptance. Should full access to data be required for peer review, authors must provide it. *Nature Communications* encourage provision of other source data in unstructured public depositories such as Dryad or figshare, or as supplementary information. To maximize data reuse, *Nature Communications* encourage publication of detailed descriptions of datasets in Scientific Data.

## Code Availability

Research papers using custom computer code will also be asked to fill out a code and software submission checklist that will be made available to editors and reviewers during manuscript assessment. The aim is to make studies that use such code more reliable by ensuring that all relevant documentation is available and by facilitating testing of software by the reviewers. Further detailed guidance and required documentation at submission and acceptance of the manuscript can be found here.

## Acknowledgements

This work was supported by the National Natural Science Foundation of China (Grant No.61873307), the Hebei Natural Science Foundation (Grant No. This work was supported by the National Natural Science Foundation of China (Grant No.61873307), the Hebei Natural Science Foundation (Grant No. F2020501040, F2021203070, F2021501021), the Fundamental Research Funds for the Central Universities under Grant N2123004, the Administration of Central Funds Guiding the Local Science and Technology Development (Grant No. 206Z1702G) and in part by the Chinese Academy of Sciences (CAS)-Research Grants Council (RGC) Joint Laboratory Funding Scheme under Project JLFS/E-104/18. (Corresponding author: Y. Zhao and Wen J. Li.).

## Ethics declarations

### Competing interests

Submission of a competing interests statement is required for all content of the journal.

# Supplementary Information

Supplementary Information: should be combined and supplied as a separate file, preferably in PDF format.

Supplementary information includes:

- material that is essential background, but which is too large, impractical or specialised to justify inclusion in the PDF version of the paper.
- tables larger than one page (in general, >50 rows or >10 columns) should be provided as tabular data files rather than a PDF.